\documentclass[10pt,aps,prd,twocolumn,showpacs,showkeys,superscriptaddress,eqsecnum,nofootinbib,longbibliography]{revtex4-1}

\makeatletter
\renewcommand\paragraph{\@startsection{paragraph}{4}{\z@}%
    {2.0ex \@plus1ex \@minus.2ex}
    {0.5ex}
    {\normalfont\normalsize}}
\makeatother
\usepackage{amsmath}        
\usepackage{amssymb}        
\usepackage{amsfonts}       
\usepackage{amsthm}         
\usepackage{mathrsfs}       
\usepackage{latexsym}       
\usepackage{graphicx}       
\usepackage{txfonts}        
\usepackage{cancel}         
\usepackage{appendix}       
\usepackage{array}          
\usepackage{booktabs}       
\usepackage{dcolumn}        
\usepackage{enumitem}       
\usepackage{placeins}
\usepackage{subcaption}     
\usepackage{mathtools}      
\usepackage{relsize}        
\usepackage{bm}             
\usepackage[table]{xcolor}         

\usepackage[
    bookmarks=true,
    colorlinks=true,
    linkcolor=blue,
    citecolor=blue,
    urlcolor=blue,
    breaklinks=true,
    pdfborder={0 0 0}
]{hyperref}

\allowdisplaybreaks

\AtBeginDocument{
  \setlength{\parskip}{0.3em}
}

\newcolumntype{d}[1]{D{.}{.}{#1}}
\newcolumntype{s}[1]{D{(}{)}{#1}}  

\begin{document}

\title{Joint Model-independent Constraints on the Post-Newtonian Parameter and Cosmic Curvature from Galaxy-scale Strong Lensing, SNe Ia, BAO, and Cosmic Chronometers}

\author{Fu-you Zheng}
\email{fyzhengdw@163.com}
\affiliation{School of Microelectronics and Artificial Intelligence, Kaili University, Kaili 556011, People's Republic of China}

\author{Xiao-jun Gao}
\email{gaoxiaojun@gznu.edu.cn (corresponding author)}
\affiliation{School of Physics and Electronics Science, Guizhou Normal University, Guiyang 550025, People's Republic of China}

\author{Hou-bing Tang}
\affiliation{School of Microelectronics and Artificial Intelligence, Kaili University, Kaili 556011, People's Republic of China}

\author{Wei Xiang}
\affiliation{School of Microelectronics and Artificial Intelligence, Kaili University, Kaili 556011, People's Republic of China}

\begin{abstract}
This investigation aims to test the validity of general relativity (GR) on kiloparsec scales via galaxy-scale strong gravitational lensing (SGL) combined with multiple cosmological probes. To circumvent the circularity problem induced by the presumption of a cosmological model based on GR, we apply the distance sum rule in the Friedmann-Lema\^{\i}tre-Robertson-Walker (FLRW) metric to constrain the post-Newtonian parameter $\gamma_{\rm PPN}$ and spatial curvature $\Omega_k$ independently of any cosmological model. The study  introduces two methodological improvements: (i) parameterizing the dark‑energy evolution factor with second‑order Chebyshev orthogonal polynomials, which ensures uniformly high convergence across the full redshift baseline while retaining $\Omega_k$ as an explicit free parameter; (ii) extending the original two‑probe configuration (SGL + Type Ia supernovae) to a four‑probe joint analysis that includes baryon acoustic oscillations and cosmic chronometers. Our results show that $\gamma_{\rm PPN} = 1.119_{-0.072}^{+0.072}$ and $\Omega_k = 0.101_{-0.076}^{+0.077}$ at 68\% confidence. The $\gamma_{\rm PPN}$ value is consistent with the GR prediction of unity at 1.7$\sigma$, while $\Omega_k$ favors a mildly open universe, with flatness consistent at $2\sigma$. Taking model-independent constraints as a benchmark, we quantify the systematic shift in $\gamma_{\rm PPN}$ induced by eight representative dark
energy models, and find that all offsets are negligible, well below the baseline $1\sigma$ uncertainty, with no significant bias at current precision. These results confirm that conventional strong-lensing gravity tests introduce no measurable systematic bias, validating our model-independent framework.
\end{abstract}


\maketitle

\section{Introduction}
\label{sec:intro}
For over a century, Einstein’s general relativity (GR) has been a cornerstone of modern physics, and testing its validity across cosmic scales remains a central endeavor in gravitational physics \cite{will2006living,will2014confrontation}. At the post-Newtonian level, the parameterized post-Newtonian (PPN) parameter $\gamma_{\rm PPN}$—which quantifies the spatial curvature produced per unit rest mass—provides the most direct probe of the weak-field metric, with GR predicting $\gamma_{\rm PPN} \equiv 1$ \cite{will2006living,will2014confrontation}. On solar system scales, the Cassini spacecraft measured $\gamma_{\rm PPN} = 1 + (2.1 \pm 2.3) \times 10^{-5}$
via the Shapiro time delay, confirming GR to a precision of $10^{-3}\%$ \cite{bertotti2003test}. By contrast, on extragalactic and cosmological scales, constraints remain orders of magnitude less precise, leaving ample room for the scale-dependent deviations predicted by modified gravity theories proposed to account for cosmic acceleration \cite{will2006living,will2014confrontation}. The parameterized post-Newtonian cosmology (PPNC) framework extends the local PPN formalism to cosmological scales in a self-consistent manner, providing a unified parameterization of gravitational couplings across scales and a theoretical basis for cross-scale tests of gravity \cite{sanghai2017parameterized, thomas2023scale}.

Galaxy-scale strong gravitational lensing (SGL) combined with stellar kinematics of lens galaxies provides a powerful tool for testing GR on kiloparsec scales \cite{schneider1992gravitational,cao2015lensing}. Comparing the lensing mass inferred from the Einstein angle with the dynamical mass from stellar velocity dispersions yields direct constraints on $\gamma_{\rm PPN}$ \cite{cao2017strong,liu2022grlensing}. However, a persistent limitation of conventional SGL-based tests is inherent circularity: the angular diameter distances needed to convert lensing observables into mass estimates are computed within the flat $\Lambda$CDM framework, which itself relies on GR. In effect, GR is tested using distance priors derived assuming GR, introducing an intrinsic bias into the inferred constraints \cite{cao2017strong,liu2022grlensing}. For instance, Cao et al. (2017) \cite{cao2017strong} constrained the PPN parameter under flat $\Lambda$CDM and, via simulations, identified a significant degeneracy between 
$\gamma_{\rm PPN}$ and spatial curvature $\Omega_k$, whereas Liu et al. (2022) \cite{liu2022grlensing} used 161 lenses with Gaussian-process distance calibration to circumvent the GR-based cosmological prior. However, such calibration-based approaches still rely on empirical statistical priors rather than fundamental geometric relations, and a fully model-independent resolution lies in the intrinsic structure of the Friedmann-Lema\^{\i}tre-Robertson-Walker (FLRW) metric.

The distance sum rule (DSR) in the FLRW metric circumvents this circularity by relating the three angular diameter distances (i.e., the distances from the observer to
the lens, $D_l$, the observer to the source, $D_s$, and the lens to the
source, $D_{ls}$) solely through $\Omega_k$. Namely, one can directly obtain the two distances $D_l$ and $D_s$ from observations of Type Ia supernovae (SNe Ia), but not the distance $D_{ls}$. The DSR can provide a relation that expresses $D_{ls}$ via $D_l$, $D_s$ and $\Omega_{k}$. Wei et al. (2022) \cite{wei2022ppn} first implemented this DSR-based framework with observational data, combining 120 strong-lensing systems with the SNe Ia sample to obtain $\gamma_{\rm PPN} = 1.11_{-0.09}^{+0.11}$ and $\Omega_k = 0.48_{-0.71}^{+1.09}$ at 68\% confidence, independently of any specific cosmological model.

Although this pioneering study by Wei et al. \cite{wei2022ppn} established the DSR-based testing paradigm and first demonstrated that $\gamma_{\rm PPN}$ and $\Omega_k$ can be constrained simultaneously without assuming any specific cosmological model, it should be emphasized that some drawbacks exist in the specific treatment methodology. For example, the dimensionless comoving distance $d(z)$ is parameterized as a third-order polynomial in redshift $z$, which exhibits degraded convergence at the high-redshift end and yields non-uniform approximation accuracy across the full redshift range; moreover, from a theoretical perspective, direct polynomial expansion of $d(z)$ implicitly encodes curvature in its coefficients, while $\Omega_k$
enters explicitly via the DSR. This double counting of curvature dilutes the geometric meaning of 
$\Omega_k$ and risks double counting in the inference. To overcome these problems, we replace the direct redshift-polynomial parameterization with a Chebyshev orthogonal polynomial expansion of the dark-energy evolution factor. This methodology preserves $\Omega_k$ as an explicit, purely geometric free parameter throughout the analysis, with curvature entering the formalism exclusively through the DSR. 
In addition, we extend the original two‑probe (SGL + SNe Ia) framework with baryon acoustic oscillations (BAO) and cosmic chronometers (CC) as complementary probes, and employ the DSR in the FLRW metric to obtain cosmology‑independent constraints on $\gamma_{\rm PPN}$ and $\Omega_k$. The resulting four‑probe joint analysis substantially breaks the strong $\gamma_{\rm PPN}$–$\Omega_k$ degeneracy that dominated the two‑probe uncertainty budget and markedly improves constraints on both parameters.

The degeneracy among dynamical dark-energy models—where different equation-of-state parameterizations yield nearly identical distance–redshift relations over the redshift range of geometric probes \cite{huterer2001degen,corasaniti2003model,giare2025overview,scherrer2015mapping}—has long been attributed to limited statistical precision, with the expectation that higher-precision data would resolve model degeneracies \cite{huterer2001degen,scherrer2015mapping}. Yet whether it arises instead from a fundamental geometric information limit in low-redshift distance measurements, rather than a statistical limitation, remains unclear, and its implications for systematic biases in extragalactic gravity tests remain unquantified. To interpret the observed bias pattern, we quantify systematic shifts in $\gamma_{\rm PPN}$ across eight representative dark-energy models. A term-by-term Taylor expansion of the dark-energy evolution factor in the compressed redshift variable 
$u = z/(1+z)$ shows that the similarity of low-redshift constraints stems from an intrinsic geometric degeneracy: model-discriminating dynamical information resides predominantly in expansion orders beyond the effective reach of geometric probes at 
$z \lesssim 2.3$.

This paper is organized as follows: in Section \ref{sec:framework}, we introduce the methodology and observations used to test GR on galactic scales. In Section \ref{sec:results}, we derive cosmology‑independent and dark‑energy‑dependent constraints on $\gamma_{\rm PPN}$ and $\Omega_k$, respectively, and quantitatively compare the biases of eight dark‑energy models with those of the cosmology‑independent case. We further interpret the physical origin of the observed bias pattern in terms of order‑by‑order geometric degeneracy, discuss the broader implications for the field, and outline the limitations of this work. We
make a summary in Section \ref{sec:conclusion}.

\section{Methodology and Data}
\label{sec:framework}
In this section, we introduce the methodology and data for constraints on both $\gamma_{\rm PPN}$ and $\Omega_k$. Particularly, relative to the study of Wei et al. (2022) \cite{wei2022ppn}, we parameterize the dark-energy evolution factor using second-order Chebyshev
orthogonal polynomials and extend the original two‑probe configuration (SGL + SNe Ia) to a four‑probe joint framework by incorporating BAO and CC.

\subsection{Strong Gravitational Lensing}
\label{subsec:lensing_ppn}

In the weak gravitational field approximation, the spacetime metric around a central object with the mass $M$ can be expressed by the PPN parameter $\gamma_{\rm PPN}$ \cite{will2006living,will2014confrontation,thorne1971foundations}, and its general form is given by \cite{wei2022ppn}
\begin{equation}
ds^2 = c^2 \left(1 - \frac{2GM}{c^2 r}\right) dt^2 - \left(1 + 2\gamma_{\rm PPN} \frac{GM}{c^2 r}\right) dr^2 - r^2 d\Omega^2,
\label{eq:ppn_metric}
\end{equation}
where $c$ is the speed of light \footnote{In this paper, we adopt $c=3\times10^5\ \mathrm{km\, s^{-1}}$ in numerical calculations.}, $G$ denotes the Newtonian gravitational constant, and $d\Omega^2 \equiv d\theta^2 + \sin^2\theta d\phi^2$ represents the solid angle element. Note that the $\gamma_{\rm PPN}=1$ for GR.

In order to test gravity with SGL systems, both the gravitational mass $M_E^{\rm grl}$ and the dynamical mass $M_E^{\rm dyn}$ enclosed within the Einstein ring should be equivalent as follows \cite{wei2022ppn}
\begin{equation}
M_E^{\rm grl} = M_E^{\rm dyn}.
\label{eq:mass_equality}
\end{equation}
From the theory of gravitational lensing, the relation between the gravitational mass $M_E^{\rm grl}$ and the Einstein angle $\theta_E$ \footnote{In this paper, all angular quantities used in lensing calculations, such as the Einstein angle $\theta_E$, the effective angular
half-light radius $\theta_{\rm eff}$, and the spectroscopic aperture radius
$\theta_{\rm ap}$, are converted from units of arcseconds to radians via multiplying by the conversion factor
$4.8481\times10^{-6}\ \text{rad arcsec}^{-1}$ in numerical
computations.} (reflecting the angular separations between multiple images) is given by
\cite{schneider1992gravitational,cao2015lensing,wei2022ppn}
\begin{equation}
\theta_E = \sqrt{\frac{1+\gamma_{\rm PPN}}{2}} \left( \frac{4 G M_E^{\rm grl}}{c^2} \frac{D_{ls}}{D_s D_l} \right)^{1/2},
\label{eq:einstein_angle}
\end{equation}
where $D_l$ and $D_s$ denote the angular diameter distance from the observer to the lens and to the source, respectively, and $D_{ls}$ is the angular diameter distance from the lens to the source. In terms of the Einstein ring radius $R_E = \theta_E D_l$, the Eq. (\ref{eq:einstein_angle}) is rewritten as
\begin{equation}
\frac{G M_E^{\rm grl}}{R_E} = \frac{2}{1+\gamma_{\rm PPN}} \cdot \frac{c^2}{4} \cdot \frac{D_s}{D_{ls}} \theta_E.
\label{eq:gm_over_re}
\end{equation}

To infer the dynamical mass $M_E^{\rm dyn}$ enclosed within the Einstein ring from the spectroscopic measurements of the lensing galaxy velocity dispersion $\sigma$, here we adopt a general mass model with power-law density profiles to describe
the mass distribution of lens galaxy
\cite{wei2022ppn,koopmans2006sloan,chen2019strong}
\begin{equation}
\begin{cases}
\rho(r) = \rho_0 \left( r / r_0 \right)^{-\alpha}, \\
\nu(r) = \nu_0 \left( r / r_0 \right)^{-\delta}, \\
\beta(r) = 1 - \sigma_t^2 / \sigma_r^2,
\end{cases}
\label{eq:power_law_profiles}
\end{equation}
where $r$ is the spherical radial coordinate originating from the center of the lensing galaxy; $\rho(r)$ represents the total (luminous baryonic plus dark matter) mass density at radius $r$, $\rho_0$ is the normalization constant of the total mass density (i.e., the total mass density at the characteristic scale radius $r = r_0$), $r_0$ denotes the characteristic scale radius serving as the unified normalization reference for both the total mass and stellar luminosity density profiles; $\alpha$ is the power-law slope of the total mass density profile; $\nu(r)$ denotes the stellar luminosity density at radius $r$, $\nu_0$ is the normalization constant of the stellar luminosity density (i.e., the stellar luminosity density at the characteristic scale radius $r = r_0$); $\delta$ is the power-law slope of the stellar luminosity density profile; $\beta(r)$ represents the anisotropy parameter of the stellar velocity dispersion, and $\sigma_t^2$ and $\sigma_r^2$ are the tangential and radial components of the stellar velocity dispersion, respectively. To account for the structural evolution of early-type lensing galaxies, one further parameterize the total mass density slope $\alpha$ as a function of the lens redshift $z_l$ and the normalized surface mass density $\tilde{\Sigma}$ \cite{wei2022ppn,chen2019strong}
\begin{equation}
\alpha = \alpha_0 + \alpha_z z_l + \alpha_s \log_{10} \tilde{\Sigma},
\label{eq:alpha_param}
\end{equation}
with $\tilde{\Sigma}$ computed as
\begin{equation}
\tilde{\Sigma} = \frac{(\sigma_0/100\ \mathrm{km\, s^{-1}})^2}{R_{\rm eff}/(10\ h^{-1}\ \mathrm{kpc})},
\label{eq:sigma_tilde}
\end{equation}
where $\sigma_0$ denotes the observed velocity dispersion; $R_{\rm eff}$ is the half-light radius of the lensing galaxy, given by $R_{\rm eff}=10^{3}\times\theta_{\rm eff}\times(c/H_0)\times d_l/(1+z_l)$ (in kpc), where $\theta_{\rm eff}$ is the effective angular radius of the
lensing galaxy, $H_0$ is the Hubble constant, and $d_l \equiv d(0,z_l)$ denotes the dimensionless comoving distance to the lens at redshift $z_l$;
$h = H_0/(100\ \mathrm{km\, s^{-1}\ Mpc^{-1}})$ denotes the reduced Hubble constant.

Based on the well-known radial Jeans equation in spherical coordinates, the radial velocity dispersion of luminous matter in early-type lens galaxies could be expressed as \cite{wei2022ppn}
\begin{equation}
\sigma_{r}^{2}(r)=\frac{G \int_{r}^{\infty} d r' r^{\prime 2 \beta-2} \nu\left(r'\right) M\left(r'\right)}{r^{2 \beta} \nu(r)},
\label{eq:jeans_equation}
\end{equation}
where $M(r)$ is the total mass contained within a spherical radius $r$. From the mass density profiles in Eq. \eqref{eq:power_law_profiles}, the relation between the dynamical mass $M_E^{\rm dyn}$ enclosed within the Einstein ring radius $R_E$ and $M(r)$ is given by \cite{wei2022ppn,koopmans2006sloan,chen2019strong}
\begin{equation}
M(r)=\frac{2}{\sqrt{\pi}} \frac{1}{\lambda(\alpha)}\left(\frac{r}{R_{E}}\right)^{3-\alpha} M_{E}^{\rm dyn},
\label{eq:mass_radius_relation}
\end{equation}
where $\lambda(x)=\Gamma\left[(x-1)/2\right] / \Gamma(x/2)$ \footnote{The Gamma function is defined via the Euler integral of the second kind:
$\Gamma(x) = \int_{0}^{+\infty} t^{x-1} e^{-t} dt$ which converges for positive real arguments $x > 0$. Standard special values include $\Gamma(1) = 1$ and $\Gamma\left(\frac{1}{2}\right) = \sqrt{\pi}$.} represents the ratio of two respective Gamma functions. Substituting Eqs. \eqref{eq:mass_radius_relation} and \eqref{eq:power_law_profiles} into Eq. \eqref{eq:jeans_equation} yields
\begin{equation}
\sigma_{r}^{2}(r)=\frac{2}{\sqrt{\pi}} \frac{G M_{E}^{\rm dyn}}{R_{E}} \frac{1}{\xi-2 \beta} \frac{1}{\lambda(\alpha)}\left(\frac{r}{R_{E}}\right)^{2-\alpha},
\label{eq:radial_dispersion}
\end{equation}
with $\xi=\alpha+\delta-2$.

The actual measurement of velocity dispersion of the lensing galaxy is the component of luminosity weighted
along the line of sight and over the effective spectrometer aperture $R_A$. For $r \leq R_A$, the Eq. \eqref{eq:radial_dispersion} can be rewritten as (see \cite{wei2022ppn,chen2019strong} for detailed derivation)
\begin{equation}
\sigma_0^2 = \frac{2}{\sqrt{\pi}} \frac{G M_E^{\rm dyn}}{R_E} F(\alpha, \beta, \delta) \left( \frac{R_A}{R_E} \right)^{2-\alpha},
\label{eq:sigma0_aperture}
\end{equation}
where the $F$ function is defined as \cite{wei2022ppn,chen2019strong}
\begin{equation}
F(\alpha, \beta, \delta) = \frac{(3-\delta)\left[\lambda(\alpha+\delta-2) - \beta\lambda(\alpha+\delta)\right]}{(\alpha+\delta-2-2\beta)(5-\alpha-\delta)\lambda(\alpha)\lambda(\delta)},
\label{eq:F_func}
\end{equation}
with $\lambda(x)$ being the ratio of Gamma functions.
The observed velocity dispersion $\sigma_{\rm ap}$ (measured within aperture $\theta_{\rm ap}$) is corrected to the standard aperture $\theta_{\rm eff}/2$ \cite{wei2022ppn,chen2019strong}
\begin{equation}
\sigma_0^{\rm obs} = \sigma_{\rm ap} \left( \frac{\theta_{\rm eff}}{2\theta_{\rm ap}} \right)^\eta,
\label{eq:sigma_obs}
\end{equation}
with aperture correction parameter $\eta=-0.066\pm0.035$ from \cite{cappellari2006sauron}. The total uncertainty of the aperture-corrected velocity dispersion, including statistical, aperture correction, and systematic components, is expressed as \cite{wei2022ppn,chen2019strong}
\begin{equation}
\left( \Delta \sigma_0^{\rm tot} \right)^2 = \left( \Delta \sigma_0^{\rm stat} \right)^2 + \left( \Delta \sigma_0^{\rm AC} \right)^2 + \left( \Delta \sigma_0^{\rm sys} \right)^2,
\label{eq:sigma0_total_uncertainty}
\end{equation}
where $ \Delta \sigma_0^{\rm stat}$ denotes the statistical uncertainty propagated from the measurement error of $\sigma_{ap}$, $\Delta \sigma_0^{\rm AC}$ is caused by the uncertainty in $\eta$. The systematic uncertainty arising from extra mass contributions of other matter along the line of sight can be quantified as
$\Delta \sigma_0^{\rm sys}$, and this effect contributes an uncertainty of approximately 3\% to the velocity dispersion \cite{liu2022grlensing,wei2022ppn,jiang2007mass}.

Based on the Eq.(\ref{eq:gm_over_re}) and the relation $R_A/R_E=(\theta_{eff}/2)/\theta_E$, the Eq. (\ref{eq:sigma0_aperture}) is rewritten as \cite{koopmans2006sloan,wei2022ppn}
\begin{equation}
\sigma_0^{\rm th} = \sqrt{ \frac{c^2}{2\sqrt{\pi}} \cdot \frac{2}{1+\gamma_{\rm PPN}} \cdot \frac{D_s}{D_{ls}} \theta_E F(\alpha, \beta, \delta) \left( \frac{\theta_{\rm eff}}{2\theta_E} \right)^{2-\alpha} }.
\label{eq:sigma_th}
\end{equation}

\subsubsection{Marginalization of Stellar Dynamical Nuisance Parameters}
\label{subsubsec:nuisance_marg}

Our SGL sample does not provide per-system measurements of the
three-dimensional luminosity density slope $\delta$ and the velocity
anisotropy parameter $\beta$. To prevent an unnecessary expansion of
the MCMC parameter space, we incorporate $\delta$ and $\beta$ as
population-level nuisance parameters, in accordance with standard
strong-lensing methodology \cite{chen2019strong,schwab2010lensing}. We perform full Bayesian marginalization over both parameters via two-dimensional numerical integration of their joint prior distribution during likelihood evaluation, i.e., neither $\delta$ nor $\beta$ is included in the set of free parameters sampled by our MCMC chains. Each lens system is marginalized independently against the same
population prior, and the resulting marginal likelihoods are multiplied
across systems.

For $\delta$, we adopt a global Gaussian prior centered on the
population-averaged value from the 130 early-type  SGL systems of Chen
et al. (2019) \cite{chen2019strong}, where the dispersion term accounts
for the intrinsic scatter of massive early-type lenses:
\begin{equation}
\delta \sim \mathcal{N}\left(\mu_\delta = 2.173,\ \sigma_\delta = 0.085\right).
\label{eq:delta_prior}
\end{equation}
For $\beta$, we adopt the canonical Gaussian prior
$\beta=0.18\pm0.13$, established for nearby early-type galaxies by
Gerhard et al. (2001) and explicitly parameterized in this Gaussian
form by Schwab et al. (2010)
\cite{schwab2010lensing,gerhard2001dynamical}; this prior is also
adopted by Chen et al. (2019) and Wei et al. (2022) for the
anisotropy marginalization \cite{wei2022ppn,chen2019strong}:
\begin{equation}
\beta \sim \mathcal{N}\left(\mu_\beta = 0.18,\ \sigma_\beta = 0.13\right).
\label{eq:beta_prior}
\end{equation}
Here the quoted widths represent the intrinsic population scatter
rather than uncertainties on the means. In this work, we assume $\delta$
and $\beta$ to be statistically independent, such that their joint prior
distribution is the product of their respective Gaussian priors. Numerically, both priors are truncated at $\pm3\sigma$, renormalized over
the truncated interval, and integrated on a $60\times60$ trapezoidal grid.

\subsubsection{SGL Observational Data}
\label{subsubsec:sgl_obs_data}
Our galaxy‑scale SGL sample is derived from the catalog of 161 systems presented by Chen et al. \cite{chen2019strong}, combining gravitational lensing geometry with stellar velocity dispersion data from spectroscopic observations of early‑type lens galaxies.

We impose a redshift cut to align the SGL sample with the Pantheon SNe Ia baseline, which covers $0.01 < z < 2.3$ and provides the distance calibration for our joint analysis. To avoid systematics from extrapolating the distance relation beyond this redshift range, we restrict the SGL sample to lens systems with source redshifts $z_s < 2.3$.

After applying this cut, we retain a final sample of 135 well‑characterized SGL systems, with lens and source redshifts spanning $0.0625 \leq z_l \leq 0.884$ and 
$0.197 \leq z_s \leq 2.2649$, respectively. For each system, the key observables are the lens redshift $z_l$, source redshift $z_s$, Einstein angle 
$\theta_E$, effective angular radius $\theta_{\rm eff}$, spectroscopic aperture radius $\theta_{\rm ap}$ (all in arcseconds), and the aperture‑averaged stellar velocity dispersion
$\sigma_{\rm ap}$ (in $\mathrm{km\,s^{-1}}$ ).

\subsection{Distance Sum Rule}
\label{subsec:distance_sum_rule}

The constraint on $\gamma_{\rm PPN}$ relies critically on the lensing distance ratio $D_s/D_{ls}$, which links the observed velocity dispersion to the theoretical prediction in Eq. (\ref{eq:sigma_th}). Conventionally, this ratio is computed assuming a standard $\Lambda$CDM cosmological model, which results in a circular reasoning problem. Namely, the cosmological model itself is built upon GR. In order to avoid the circularity problem, following the reference \cite{wei2022ppn}, we adopt the DSR along null geodesics of the FLRW metric in a homogeneous and isotropic universe to determine the lensing distance ratio in a cosmology-independent way, and the corresponding dimensionless comoving distance $d(z_{l}, z_{s}) \equiv(H_{0} / c)(1+z_{s}) D_{A}(z_{l}, z_{s})$ is expressed by \cite{wei2022ppn}
\begin{equation}
d\left(z_{l}, z_{s}\right)=\frac{1}{\sqrt{\left|\Omega_{k}\right|}} \text{sinn}\left( \sqrt{\left|\Omega_{k}\right|} \int_{z_{l}}^{z_{s}} \frac{d z'}{E\left(z'\right)} \right),
\label{eq:comoving_distance_flrw}
\end{equation}
where $\Omega_k$ represents the spatial curvature of the universe and $E(z)\equiv H(z)/H_0$ is the dimensionless Hubble parameter.
$\mathrm{sinn}(x) = \sinh(x)$ for an open universe ($\Omega_k > 0$), and $\mathrm{sinn}(x) = \sin(x)$ for a closed universe ($\Omega_k < 0$). The Eq. \eqref{eq:comoving_distance_flrw} reduces to a linear function of the integral for a flat universe with $\Omega_k=0$.
For convenience of notation, we define the following dimensionless comoving distances: $d(z) \equiv d(0, z)$ (from the observer at redshift 0 to an object at redshift $z$), $d_l \equiv d(0, z_l)$ (from the observer to the lensing galaxy at redshift $z_l$), $d_s \equiv d(0, z_s)$ (from the observer to the background source at redshift $z_s$), and $d_{ls} \equiv d(z_l, z_s)$ (from the lensing galaxy to the background source). 

Then, one can derive a simple form of the DSR in the FLRW framework as follows \cite{wei2022ppn,peebles1993principles,bernstein2006dark,rasanen2015direct}

\begin{equation}
\frac{d_{ls}}{d_s} = \sqrt{1+\Omega_k d_l^2} - \frac{d_l}{d_s} \sqrt{1+\Omega_k d_s^2}.
\label{eq:distance_sum_rule}
\end{equation}
This relation is very general because it only assumes that
geometrical optics holds and that light propagation is described
by the FLRW metric.

The $d_l$ and $d_s$ on the right side of Eq. (\ref{eq:distance_sum_rule}) can be independently measured, and then the dimensionless distance ratio $d_{ls}/d_s$\footnote{Note that $d_{ls}/d_s$ is equivalent to the ratio of the angular diameter distances $D_{ls}/D_s$.} (depending only on the curvature parameter) can be evaluated. Lastly, we can obtain direct constraints on both $\gamma_{\rm PPN}$ and $\Omega_k$ using Eqs. (\ref{eq:sigma_th}) and (\ref{eq:distance_sum_rule}), without assuming a particular cosmological model.

Multiple approaches have been proposed to reconstruct the dimensionless comoving distance $d(z)$. Early parametric schemes directly parameterize $d(z)$ as a redshift polynomial. These have been applied to Hubble‑constant determinations using time‑delay lensing \cite{qi2020hubble}, as well as to the first joint observational constraints on the post‑Newtonian parameter $\gamma_{\rm PPN}$ and spatial curvature $\Omega_k$ from galaxy‑scale strong gravitational lensing \cite{wei2022ppn}. Subsequent nonparametric alternatives include Gaussian‑process regression \cite{liu2022grlensing} and fully data‑driven artificial‑neural‑network reconstructions \cite{ran2024extragalactic}, which relax assumptions about the functional form.

Direct parameterization of $d(z)$ as a redshift polynomial (i.e. $d(z) = z + a_1 z^2 + a_2 z^3$, where $a_1$ and $a_2$ are polynomial coefficients.) exhibits poor convergence at high redshifts and yields non‑uniform approximation accuracy over the full redshift range. The polynomial coefficients implicitly encode curvature information, whereas $\Omega_k$ appears explicitly in the formalism through the DSR. This double‑counting of curvature contaminates the pure geometric interpretation of $\Omega_k$ and may introduce biases from redundant counting in parameter inference. Furthermore, Gaussian‑process regression and neural‑network‑based reconstructions generally suffer from limited physical interpretability. None of these existing frameworks can explicitly disentangle the separate contributions of matter, spatial curvature, and dark energy to the distance‑redshift relation.

To overcome these limitations, we adopt a parameterization of the dark‑energy evolution factor built upon second‑order Chebyshev orthogonal polynomials. This formulation treats $\Omega_k$ as an explicit purely geometric free parameter and yields uniform high‑fidelity approximations across the entire redshift range. Further details of this parameterization are provided in the following subsection.

\subsection{Chebyshev parameterization of the Dark‑Energy Evolution Factor}
\label{subsec:chebyshev_param}

The dimensionless Hubble parameter $E(z) \equiv H(z)/H_0$ follows from the Friedmann equations in the homogeneous and isotropic FLRW metric. For the late universe with negligible radiation, its general form for a universe composed of non-relativistic matter, dark energy, and spatial curvature is given by \cite{magana2017testing}
\begin{equation}
E^2(z)=\Omega_m(1+z)^3+\Omega_k(1+z)^2+\Omega_{\rm de} f_{\rm de}(z),
\label{eq:E2_general1}
\end{equation}
\begin{equation}
f_{\rm de}(z) \equiv \frac{\Omega_{\rm de}(z)}{\Omega_{\rm de}}=e^{3\int^z_0\dfrac{1+w(z')}{1+z'}dz'},
\label{eq:E2_general2}
\end{equation}
where $\Omega_m$ is the present‑day matter density fraction; $\Omega_{\rm de}$ denotes the present‑day dark energy density fraction, satisfying the normalization $\Omega_{\rm de}\equiv1-\Omega_m-\Omega_k$; $f_{\rm de}(z)$ is the dark energy evolution factor; $\Omega_{\rm de}(z)$ denotes the dark energy density fraction at redshift $z$; $w(z)\equiv p_{\rm de}/\rho_{\rm de}$ is the redshift‑dependent equation of state of dark energy, with $p_{\rm de}$ and $\rho_{\rm de}$ its pressure and energy density, respectively.

We further expand $f_{\rm de}(z)$ in second‑order Chebyshev orthogonal polynomials, first performing a redshift variable transformation
\begin{equation}
u = \frac{z}{1+z}, \quad t = 2\frac{u}{u_{\rm max}} - 1,
\end{equation}
which maps the observed redshift interval $z \in [0, z_{\rm max}]$ (with $z_{\rm max}=2.33$ in this work) onto the canonical domain $t \in [-1, 1]$ of the first‑kind Chebyshev polynomials $T_n(t)$. The dark energy evolution factor is then expanded as
\begin{equation}
f_{\rm de}(z) = 1 + c_1 \left[T_1(t) + 1\right] + c_2 \left[T_2(t) - 1\right],
\label{eq:fde_chebyshev}
\end{equation}
where $T_1(t)=t$ and $T_2(t)=2t^2-1$ are the first two Chebyshev polynomials of the first kind, and $c_1$ and $c_2$ are free expansion coefficients. This construction automatically satisfies the physical boundary condition $f_{\rm de}(z=0)=1$, as required by the present‑day dark energy density normalization.

Substituting the Eq.~(\ref{eq:fde_chebyshev}) into the Eq.~(\ref{eq:E2_general1}), we obtain the parameterized form of the dimensionless Hubble parameter $E(z)$, which is independent of the specific form of the equation of state $w(z)$. The dimensionless comoving distance $d(z)$ is then computed via numerical integration of $1/E(z)$, with the spatial curvature $\Omega_k$ consistently treated as an explicit free parameter, in full agreement with the DSR formalism in Sec.~\ref{subsec:distance_sum_rule}.

In contrast to the direct redshift‑polynomial parameterization adopted in Ref.~\cite{wei2022ppn}, the Chebyshev parameterization provides three distinct methodological advantages, which we detail as follows.

\paragraph{Explicit physical meaning of parameters}
Both the matter density $\Omega_m$ and the spatial curvature $\Omega_k$ are treated as independent free parameters with well‑defined physical meanings; only the dark‑energy evolution is parameterized in a model‑agnostic manner. This effectively eliminates the parameter degeneracy inherent to the direct polynomial parameterization, thereby ensuring that the $\Omega_k$ inferred from the distance sum rule strictly corresponds to the geometric spatial curvature of the universe.

\paragraph{Uniform global convergence}
Chebyshev polynomials exhibit the minimax property, which minimizes the maximum approximation error over the entire interval and thereby ensures uniformly high precision over the entire redshift range. This circumvents the error divergence at high redshifts that plagues standard Taylor and redshift‑polynomial expansions \cite{capozziello2018chebyshev,alfano2026desi}.

\paragraph{Hierarchical structure for degeneracy analysis}
The orthogonality of the Chebyshev basis—a core mathematical property first introduced into cosmological contexts by Ref.~\cite{capozziello2018chebyshev}—guarantees the statistical independence of the expansion coefficients at different orders. This natural hierarchical structure underpins our sequential analysis of model degeneracy with respect to the compressed redshift variable $u$, as we demonstrate in Sec.~\ref{sec:discussion}.

\subsection{Type Ia Supernovae as Standard Candles for Distance Calibration}
\label{subsec:sne_theory}

SNe Ia serve as the most extensively calibrated and high-precision standard candles in observational cosmology, providing the most accurate luminosity distance measurements over a broad redshift range \cite{scolnic2018pantheon}. Their peak luminosity exhibits very small intrinsic scatter owing to the uniform Chandrasekhar mass of their progenitors at the moment of explosion; when combined with empirical corrections for light-curve shape (stretch factor) and line-of-sight dust reddening (color correction), this uniformity allows SNe Ia to act as reliable tracers of the cosmic expansion history \cite{huterer2001degen,scherrer2015mapping}. In the context of this work, SNe Ia supply the independent distance calibration required by the DSR formalism, enabling model-independent constraints on gravitational physics without prior assumptions about the nature of dark energy.

The distance modulus $\mu(z)$, defined as the difference between the observed apparent magnitude $m$ and the intrinsic absolute magnitude $M$ of a standard candle at redshift $z$, is formally related to the luminosity distance $D_L(z)$ by \cite{peebles1993principles}
\begin{equation}
\mu(z) = m - M = 5\log_{10}\left( \frac{D_L(z)}{10\,\mathrm{pc}} \right).
\label{eq:mu_def}
\end{equation}
The $10\,\mathrm{pc}$ reference distance in the denominator encodes the standard normalization of absolute magnitude; for consistency with the megaparsec units adopted universally for cosmological distances throughout this work, this reference scale is equivalent to $10^{-5}\,\mathrm{Mpc}$.

In the homogeneous and isotropic FLRW framework, the luminosity distance can be expressed in terms of the dimensionless comoving distance $d(z)$ defined in Eq.~\eqref{eq:comoving_distance_flrw}:
\begin{equation}
D_L(z) = \frac{c}{H_0} \left(1+z\right) d(z).
\label{eq:D_L_dz}
\end{equation}
This relation acts as the conversion interface that maps the SNe Ia luminosity-distance observable onto the unified comoving-distance framework adopted uniformly across all probes in this work. Substituting Eq.~\eqref{eq:D_L_dz} into Eq.~\eqref{eq:mu_def} yields the model-predicted distance modulus expressed directly in terms of the comoving coordinate:
\begin{equation}
\mu_{\rm model}(z) = 5\log_{10}\left[\left(1+z\right) d(z)\right] - 5\log_{10}\left(10\,\mathrm{pc}\cdot H_0/c\right),
\label{eq:mu_model_sne}
\end{equation}
where $d(z)$ is computed directly from the Chebyshev-parameterized dimensionless Hubble parameter in Eq.~\eqref{eq:E2_general1}. This unified distance computation pipeline ensures full numerical consistency with the comoving distance scales used for the SGL, BAO, and CC probes, eliminating relative systematic offsets between the four datasets. The absolute $B$-band peak magnitude $M_B$ enters the analysis as a global nuisance parameter responsible for the overall luminosity scale calibration of the SNe Ia sample.

\subsubsection{SNe Ia Observational Data}
\label{subsubsec:sne_obs_data}

We adopt the SNe Ia dataset from the final public Pantheon release of \cite{scolnic2018pantheon}. We apply no further processing, refitting, bias recalibration, or sample cuts to the official product.

The full sample contains 1048 spectroscopically confirmed SNe Ia in the redshift range $0.01 < z < 2.3$, drawn from 18 independent surveys. All light curves are fitted with the SALT2 spectral template and corrected for bias using the BBC method \cite{kessler2017beams} in the official reduction pipeline. The data release includes the full covariance matrix, accounting for both statistical errors and survey‑related systematics.

The original collaboration performed rigorous quality filtering on the sample, including validation of redshifts and corrected magnitudes, and applied a redshift cut at $z=0.01$ to reduce systematic effects from peculiar velocities. We adopt the full public dataset without modification, using its redshift range as the distance anchor for our joint analysis and imposing no additional cuts.

The primary observables for our cosmological fit are the CMB‑frame redshift $z$ and the bias‑corrected rest‑frame $B$‑band apparent magnitude $m_{\rm corr}$, both adopted directly from the official catalog. The total covariance matrix combines the diagonal statistical errors with the full off‑diagonal systematic covariance provided with the dataset, and its inverse is used directly in the likelihood evaluation.

\subsection{Baryon Acoustic Oscillations as a Cosmological Standard Ruler}
\label{subsec:bao_theory}

BAO arise from the propagation of sound waves through the primordial photon–baryon plasma prior to cosmic recombination. These waves imprint a characteristic comoving scale—the sound horizon at the baryon drag epoch $r_d$—onto the matter power spectrum. Within the standard framework of early‑universe microphysics, this scale is determined solely by the thermodynamics of the pre‑recombination plasma and is independent of the late‑time expansion history, spatial curvature, or modifications to gravity, thereby serving as a robust standard ruler for mapping the cosmic expansion over a broad redshift range \cite{adame2025desi}.

From the anisotropic clustering of galaxies, BAO measurements constrain two fundamental comoving distance scales independently: the transverse comoving angular diameter distance $D_M(z)$ and the line‑of‑sight Hubble distance $D_H(z) \equiv c/H(z)$. In terms of our dimensionless comoving distance $d(z)$ (defined in Eq.~\ref{eq:comoving_distance_flrw}) and the spatial curvature $\Omega_k$, their closed‑form expressions read
\begin{equation}
D_M(z) = \frac{c}{H_0} \, d(z),
\label{eq:DM_dz}
\end{equation}
\begin{equation}
D_H(z) = \frac{c}{H_0} \, \frac{dd(z)/dz}{\sqrt{1 + \Omega_k \, d(z)^2}}.
\label{eq:DH_dz}
\end{equation}

Equation~\eqref{eq:DM_dz} follows directly from the normalization that defines $d(z)$ as the dimensionless transverse comoving coordinate in Eq.~\ref{eq:comoving_distance_flrw}. To derive Eq.~\eqref{eq:DH_dz}, we start from the definition $D_H(z) \equiv c/H(z)$, take the redshift derivative of $d(z)$ as given by Eq.~\ref{eq:comoving_distance_flrw}, and combine the result with an identity from the FLRW metric that relates the Hubble parameter to the radial derivative of the comoving distance:
\begin{equation}
H(z) = H_0 \frac{\sqrt{1+\Omega_k \, d(z)^2}}{dd(z)/dz}.
\end{equation}
Substituting this identity into the definition of $D_H(z)$ yields the expression in Eq.~\eqref{eq:DH_dz}, ensuring full consistency with the comoving distance framework employed in the strong‑gravitational‑lensing analysis.

For spherically averaged BAO measurements with insufficient signal‑to‑noise ratio to resolve anisotropy, the volume‑averaged distance $D_V(z)$ is adopted as a combined distance indicator, defined as
\begin{equation}
D_V(z) = \left[ z \, D_M(z)^2 \, D_H(z) \right]^{1/3}.
\label{eq:DV_def}
\end{equation}

All BAO observables are conventionally normalized by the sound horizon $r_d$ to form the dimensionless ratios $D_M/r_d$, $D_H/r_d$, and $D_V/r_d$. Since $r_d$ is set entirely by early‑universe microphysics, it is independent of the late‑time dark energy dynamics, spatial curvature, and gravitational modifications that are the central focus of this work. We therefore adopt a fixed fiducial value calibrated from CMB observations throughout our analysis.

Specifically, we fix the sound horizon at the baryon drag epoch to $r_d = 147.09\ \mathrm{Mpc}$, corresponding to the marginalized best‑fit $\Lambda$CDM result from the final Planck 2018 data release under the TT, TE, EE + lowE + lensing likelihood \cite{planck2018_cosmo_params}. This value is calibrated by the well‑understood physics of the pre‑recombination baryon‑photon plasma. Adopting this CMB‑calibrated sound horizon anchors the relative BAO distance measurements to an absolute physical scale, following the standard inverse distance ladder formalism.

Crucially, $r_d$ is established at the baryon drag epoch ($z_d \approx 1060$), long before the late‑universe phenomena under investigation become dynamically relevant. This choice therefore does not introduce any meaningful prior bias into our tests of gravitational theory, cosmic curvature, or dark energy evolution.

The theoretical predictions for all BAO observables are computed directly from the Chebyshev‑parameterized dimensionless Hubble parameter $E(z)$ and the corresponding comoving distance $d(z)$ (Eq.~\ref{eq:E2_general1}), ensuring full numerical consistency with the distance calculations used for the SGL and SNe Ia probes.

\subsubsection{BAO Observational Data}
\label{subsubsec:bao_obs_data}
We adopt the BAO measurements from the first-year DESI DR1 cosmology analysis of \cite{adame2025desi}. As the first Stage‑IV extragalactic spectroscopic survey, DESI provides spectroscopic redshifts for over 6 million unique sources across 
$0.1 < z < 4.2$, spanning five tracer populations: the Bright Galaxy Survey (BGS), luminous red galaxies (LRGs), emission-line galaxies (ELGs), quasars (QSOs), and the Ly$\alpha$ forest.

The BAO measurements are divided into seven independent redshift bins, with observables optimized per bin according to the statistical precision of the clustering signal. For bins with sufficient signal-to-noise ratio to resolve anisotropic clustering via the Alcock–Paczyński effect (LRG1, LRG2, LRG3+ELG1, ELG2, and Ly$\alpha$ QSO), results are reported as $D_M(z)/r_d$ and 
$D_H(z)/r_d$; for the lower statistical power bins (BGS and QSO), only the spherically averaged $D_V(z)/r_d$ is provided. A complete summary of the tracer samples, their effective redshifts, and the corresponding BAO constraints are given in Table 1 of \cite{adame2025desi}.

We employ the official DESI Gaussian likelihood, whose covariance matrix is block‑diagonal, with non‑zero off‑diagonal entries only between $D_M(z)/r_d$ and $D_H(z)/r_d$ within each bin (intra‑bin correlation coefficients are provided for anisotropic measurements). Cross‑bin correlations are negligible and are therefore omitted, consistent with the default analysis of \cite{adame2025desi}. The BAO measurements were derived from a blinded pipeline, with systematic uncertainties fully characterized for theoretical modeling, the galaxy–halo connection, and observational effects.

To ensure consistent redshift coverage with our SGL and SNe Ia samples—and to minimize biases from extrapolating the distance relation beyond the Pantheon SNe Ia redshift baseline of $z<2.3$—we apply a nominal redshift cut to the DESI BAO dataset. The highest-redshift bin from the Ly$\alpha$ forest tracer has an effective redshift $z_{\rm eff}=2.330$, which lies only marginally above the $z=2.3$ baseline. Removing this bin would leave a large gap in BAO coverage above $z\simeq1.5$, eliminating constraints on the expansion history across nearly one billion years of cosmic time and substantially weakening the leverage of BAO data at the high-redshift end. After weighing the small systematic risk of mild extrapolation against the substantial loss of cosmological information from truncating the sample, we retain this highest-redshift bin in our analysis.

After this selection, the final BAO sample comprises seven bins spanning $0.295 \leq z_{\rm eff} \leq 2.330$, covering approximately 11 billion years of cosmic history and delivering robust, well‑calibrated constraints on the expansion history across the full probed redshift range.

\subsection{Cosmic Chronometers as Direct Hubble Parameter Probes}
\label{subsec:cc_theory}
CC provide a direct, model‑independent measurement of the Hubble parameter 
$H(z)$ by using the age difference between passively evolving early‑type galaxies at different redshifts. Unlike geometric distance probes such as SNe Ia and BAO, CC directly measures the expansion rate along the line of sight, thereby offering complementary constraints on the cosmic expansion history and helping to break parameter degeneracies.

The observable quantity measured by cosmic chronometers is the Hubble parameter $H(z)$. In terms of the dimensionless Hubble parameter $E(z)$ that forms the backbone of our unified parameterization framework, the relation reads:
\begin{equation}
H(z) = H_0 \, E(z).
\label{eq:H_z_E_z}
\end{equation}
In our model‑independent framework, we compute $E(z)$ directly from Eq.~\eqref{eq:E2_general1} using the Chebyshev‑parameterized dark‑energy evolution factor, with no prior assumptions on the physical nature of dark energy.

Relative to other probes, CC measurements are especially sensitive to the expansion rate (i.e., $\dot{a}/a$), yielding constraints that are orthogonal to those from the distance‑based SNe Ia and BAO datasets. This orthogonality renders them highly effective for lifting degeneracies between $\Omega_k$ and the dark‑energy expansion coefficients in our joint analysis.

In this work, we employ the DESI DR1 cosmic chronometer dataset from Loubser (2025) \cite{loubser2025}, which is derived from the ${\rm D}4000_n$
spectral index of over 360,000 massive, passively evolving galaxies ($\sigma>280\ \mathrm{km\,s^{-1}}$, $\log M_*/M_\odot>10.75$). This dataset delivers $H(z)$ measurements at three effective redshifts ($z=0.46$, $z=0.67$, $z=0.83$), with fully quantified statistical and systematic uncertainties, including highly correlated systematic errors between redshift bins.

\subsubsection{CC Observational Data}
\label{subsubsec:cc_obs_data}
Our CC sample is derived from the DESI DR1 measurements of \cite{loubser2025}, which represent the largest homogeneous CC compilation to date. The sample comprises 358,259 massive, passively evolving galaxies selected from DESI spectroscopy with rigorous criteria on stellar mass ($\log M_*/M_\odot > 10.75$), velocity dispersion ($\sigma > 280\ \mathrm{km\,s^{-1}}$), and the absence of [O II] emission, ensuring a clean sample of passive systems.

The $H(z)$ measurements are given by using the differential age method based on the ${\rm D}4000_n$ spectral index, calibrated against stellar population synthesis models. The final dataset contains three model-independent measurements of $H(z)$:
\begin{itemize}[leftmargin=0pt, label={}, itemsep=4pt, topsep=0pt, partopsep=0pt]
    \item $\begin{aligned}[t]
        H(z=0.46) &= 88.48 \pm 0.57\ (\mathrm{stat}) \pm 12.32\ (\mathrm{syst})\ \mathrm{km\,s^{-1}\,Mpc^{-1}},
    \end{aligned}$
    \item $\begin{aligned}[t]
        H(z=0.67) &= 119.45 \pm 6.39\ (\mathrm{stat}) \pm 16.64\ (\mathrm{syst})\ \mathrm{km\,s^{-1}\,Mpc^{-1}},
    \end{aligned}$
    \item $\begin{aligned}[t]
        H(z=0.83) &= 108.28 \pm 10.07\ (\mathrm{stat}) \pm 15.08\ (\mathrm{syst})\ \mathrm{km\,s^{-1}\,Mpc^{-1}}.
    \end{aligned}$
\end{itemize}

We employ the full covariance matrix from the original work, which includes both statistical uncertainties and highly correlated systematic errors arising from stellar population modeling, metallicity calibration, and the stellar library choice. The correlation coefficients among the three redshift bins are $\rho(z = 0.46, z = 0.67) = 0.932$, $\rho(z = 0.46, z = 0.83) = 0.830$, and
$\rho(z = 0.67, z = 0.83) = 0.776$; we fully propagate them into our MCMC likelihood evaluation.

\subsection{Likelihood Construction}
\label{subsec:stats}
For the SGL, we adopt the joint likelihood as the product of univariate Gaussian likelihoods for the aperture‑averaged velocity dispersion residuals, following the cosmology‑independent formalism proposed in \cite{wei2022ppn}. We marginalize this likelihood over the stellar population parameters  $\delta$ and $\beta$ using the two‑dimensional numerical integration scheme described in Section~\ref{subsubsec:nuisance_marg}, with both parameters adopting the Gaussian priors given in Eqs.~\eqref{eq:delta_prior} and \eqref{eq:beta_prior}. The full likelihood for the SGL sample is given by
\begin{equation}
\mathcal{L}_{\rm SGL} = \prod_{i=1}^{N_{\rm lens}} \frac{1}{\sqrt{2\pi}\, \Delta\sigma_{0,i}^{\rm tot}}
\exp\left[ -\frac{1}{2} \left( \frac{\sigma_{0,i}^{\rm th} - \sigma_{0,i}^{\rm obs}}{\Delta\sigma_{0,i}^{\rm tot}} \right)^{\!2} \right],
\label{eq:sgl_likelihood}
\end{equation}
where $N_{\rm lens}=135$ is the number of lensing systems that satisfy our redshift selection criteria; $\sigma_{0,i}^{\rm th}$ is the theoretical aperture-averaged velocity dispersion; $\sigma_{0,i}^{\rm obs}$ is the observed velocity dispersion; the $\Delta\sigma_{0,i}^{\rm tot}$ is the total uncertainty. We impose numerical clipping and a positive lower bound on all intermediate quantities to prevent unphysical negative values and division‑by‑zero singularities.

For the SNe Ia, we construct the likelihood under a multivariate Gaussian formalism using distance modulus residuals, employing the full Pantheon compilation \cite{scolnic2018pantheon} with its intrinsic full covariance matrix. This formulation avoids the simplifying assumption of uncorrelated diagonal errors. The negative twice log-likelihood is given by
\begin{equation}
-2\ln\mathcal{L}_{\rm SN} = \Delta{\boldsymbol{\mu}}^\mathrm{T} \cdot \mathbf{Cov}_{\rm SN}^{-1} \cdot \Delta{\boldsymbol{\mu}},
\label{eq:sn_likelihood}
\end{equation}
The superscript $^\mathrm{T}$ denotes the matrix transpose operation, with $\Delta{\boldsymbol{\mu}}^\mathrm{T}$ corresponding to the row-vector transpose of the column residual vector $\Delta{\boldsymbol{\mu}}$,
where $\Delta{\boldsymbol{\mu}} = \boldsymbol{\mu}_{\rm obs} - \boldsymbol{\mu}_{\rm model}$ denotes the residual vector quantifying the difference between observed and model-predicted distance moduli over the full supernova sample. The observed distance modulus vector $\boldsymbol{\mu}_{\rm obs}$ is constructed from the bias-corrected rest-frame B-band peak apparent magnitudes of the Pantheon sample. Its $i$-th element reads $\mu_{{\rm obs},i} = m_{{\rm corr},i} - M_B$, where $M_B$ is a global nuisance parameter for the absolute B-band peak magnitude of a fiducial SNe Ia. The model-predicted distance modulus vector $\boldsymbol{\mu}_{\rm model}$ is computed from Eq.~\eqref{eq:mu_model_sne} at the redshift of each individual SN, with the dimensionless comoving distance $d(z)$ evaluated consistently via the unified Chebyshev parameterization scheme. The total covariance matrix $\mathbf{Cov}_{\rm SN}$ of the Pantheon sample incorporates both diagonal statistical uncertainties from photometric and light-curve fitting errors and a full off-diagonal systematic covariance accounting for cross-survey calibration systematics, intrinsic luminosity scatter, and selection effects.

For the BAO measurements, we adopt the standard multivariate Gaussian likelihood pipeline from the DESI DR1 collaboration \cite{adame2025desi}. The negative twice log-likelihood corresponds to the squared Mahalanobis distance between observed and theoretical BAO scale ratios:
\begin{equation}
-2\ln\mathcal{L}_{\rm BAO} = \Delta\boldsymbol{x}_{\rm BAO}^{\mathrm{T}} \cdot \mathbf{Cov}_{\rm BAO}^{-1} \cdot \Delta\boldsymbol{x}_{\rm BAO},
\label{eq:bao_likelihood}
\end{equation}
where the residual vector is defined as $\Delta\boldsymbol{x}_{\rm BAO} = \boldsymbol{x}_{\rm BAO}^{\rm obs} - \boldsymbol{x}_{\rm BAO}^{\rm th}$. The data vector $\boldsymbol{x}_{\rm BAO}$ stacks the $D_M/r_d$, $D_H/r_d$, and $D_V/r_d$ observables in an ordering consistent with the official DESI covariance matrix. The covariance matrix $\mathbf{Cov}_{\rm BAO}$ accounts for intra-bin correlations between the transverse and line-of-sight BAO signals. We apply standard numerical regularization during matrix inversion to ensure positive definiteness and suppress unphysical distance solutions.

For the DESI DR1 cosmic chronometer Hubble parameter data, we adopt a multivariate Gaussian likelihood that incorporates correlated systematic errors across redshift bins, following the framework of \cite{loubser2025}. The likelihood reads
\begin{equation}
-2\ln\mathcal{L}_{\rm CC} = \Delta\boldsymbol{H}_{\rm CC}^{\mathrm{T}} \cdot \mathbf{Cov}_{\rm CC}^{-1} \cdot \Delta\boldsymbol{H}_{\rm CC},
\label{eq:cc_likelihood}
\end{equation}
where \(\Delta\boldsymbol{H}_{\rm CC} = \boldsymbol{H}_{\rm CC}^{\rm obs} - \boldsymbol{H}_{\rm CC}^{\rm th}\) denotes the residual vector of observed minus model-predicted \(H(z)\) values, and \(\mathbf{Cov}_{\rm CC}\) includes both uncorrelated statistical errors and fully correlated systematic contributions as tabulated in the DESI CC catalog.

We treat the four datasets—SGL, SNe Ia, BAO, and CC—as independent probes. The total joint log-likelihood is therefore the sum of the individual component log-likelihoods:
\begin{equation}
\ln\mathcal{L}_{\rm tot} = \ln\mathcal{L}_{\rm SGL} + \ln\mathcal{L}_{\rm SN} + \ln\mathcal{L}_{\rm BAO} + \ln\mathcal{L}_{\rm CC}.
\label{eq:total_loglikelihood}
\end{equation}

By Bayes' theorem, the posterior probability distribution for our full parameter vector \(\boldsymbol{\theta}\) is proportional to the product of the total likelihood and the joint prior probability \(P(\boldsymbol{\theta})\):
\begin{equation*}
\mathcal{P}(\boldsymbol{\theta} \mid \boldsymbol{D}) \propto \mathcal{L}_{\rm tot}(\boldsymbol{D} \mid \boldsymbol{\theta})\, P(\boldsymbol{\theta}).
\end{equation*}

Bayesian posterior sampling is performed using the affine-invariant ensemble sampler \texttt{emcee} \cite{foreman2013}. We follow a standard two-phase procedure: an initial burn-in phase to converge to the target distribution, followed by a production phase to collect posterior samples. Convergence is assessed via the integrated autocorrelation time, and only post-burn-in thinned samples are retained for analysis.

\section{Observational Constraints}
\label{sec:results}
In this section, we first derive the model-independent constraints on $\gamma_{\rm PPN}$, $\Omega_k$, and other background cosmological parameters. We then obtain dark-energy-dependent constraints on $\gamma_{\rm PPN}$ and $\Omega_k$, and compare the biases of eight dark-energy models against the model-independent baseline. We finally attribute the observed bias pattern to order-by-order geometric degeneracy, discuss the wider implications, and state the limitations of this study.

\subsection{Cosmology-independent Constraints on $\gamma_{\rm PPN}$ and $\Omega_k$}

\begin{figure}[htb!]
\centering
\includegraphics[width=0.9\columnwidth]{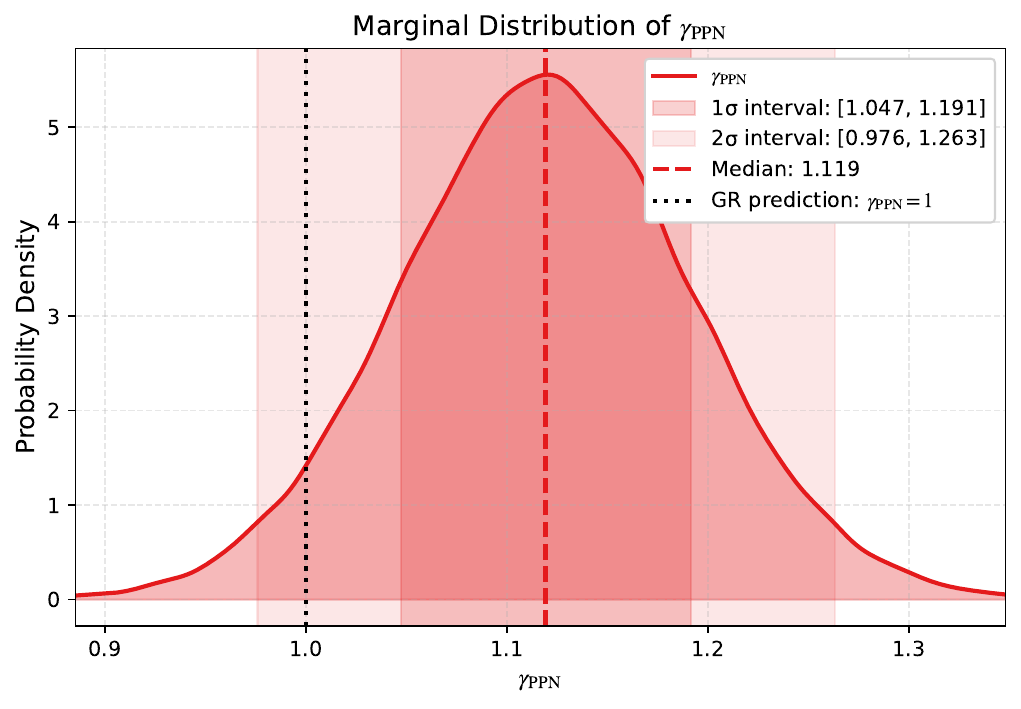}
\caption{One-dimensional marginal posterior distribution of the PPN gravity parameter $\gamma_{\rm PPN}$. The shaded regions denote the 68\% and 95\% credible intervals, the vertical dashed line marks the posterior median, and the vertical dotted line indicates the GR prediction of $\gamma_{\rm PPN} = 1$.}
\label{fig:marginal_gamma}
\end{figure}

\begin{figure}[htb!]
\centering
\includegraphics[width=0.9\columnwidth]{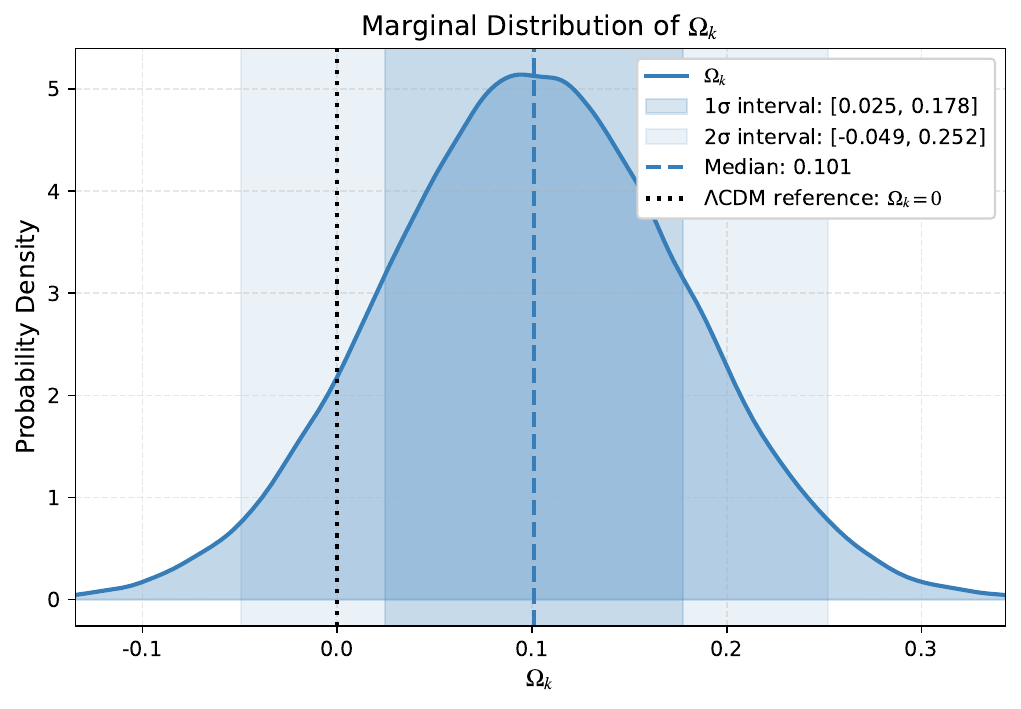}
\caption{One-dimensional marginal posterior distribution of the cosmic curvature density parameter $\Omega_k$. The shaded regions denote the 68\% and 95\% credible intervals, the vertical dashed line marks the posterior median, and the vertical dotted line indicates the flat $\Lambda$CDM prediction of $\Omega_k = 0$.}
\label{fig:marginal_omega}
\end{figure}

Figures \ref{fig:marginal_gamma} and \ref{fig:marginal_omega} show the one-dimensional marginal posterior distributions of $\gamma_{\rm PPN}$ and $\Omega_k$, respectively, obtained from the joint analysis of the SGL, SNe Ia, BAO and CC probes. The posterior for $\gamma_{\rm PPN}$ is approximately Gaussian, with a median value of $1.119_{-0.072}^{+0.072}$ (68\% C.I.: $1.047\leq\gamma_{\rm PPN}\leq1.191$). The GR prediction of $\gamma_{\rm PPN} = 1$ falls outside the 68\% credible interval but remains statistically consistent with our measurement at the $1.7\sigma$ level. For $\Omega_k$, we find a mild preference for an open universe, with a median value of $\Omega_k = 0.101_{-0.076}^{+0.077}$ (68\% C.I.: $0.025\leq\Omega_k\leq0.178$). The flat $\Lambda$CDM prediction of $\Omega_k = 0$  is located at $1.33\sigma$ below the median, placing it outside the 68\% interval but included at the $2\sigma$ level.

\begin{figure}[htb!]
\centering
\includegraphics[width=0.9\columnwidth]{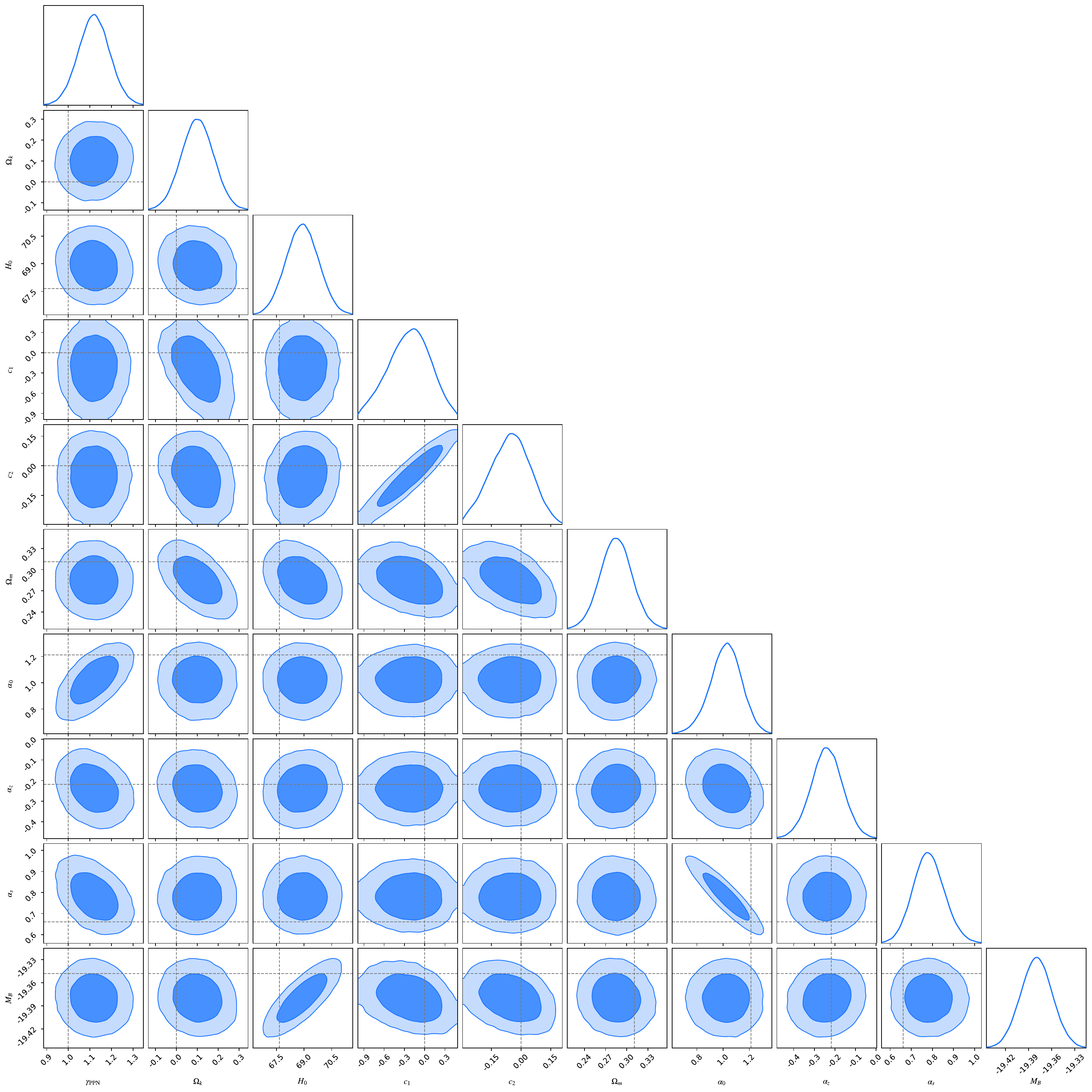}
\caption{Corner plot of the joint posterior distribution for all free parameters in our model-independent baseline analysis. Diagonal panels show one-dimensional marginalized posteriors, and off-diagonal panels present two-dimensional 68\% and 95\% credible contours. Dashed lines mark the reference values for each parameter.}
\label{fig:corner}
\end{figure}

Figure \ref{fig:corner} displays the full joint posterior distribution of all ten free parameters as a corner plot. Diagonal panels show the one-dimensional marginalized posteriors, while off-diagonal panels present the two-dimensional 68\% and 95\% credible contours. The joint contours show that adding BAO and CC probes effectively break the strong 
$\gamma_{\rm PPN}$--$\Omega_k$
degeneracy that dominated the two‑probe uncertainty budget of Wei et al. (2022) \cite{wei2022ppn}. The Chebyshev coefficients 
$c_1$ and $c_2$ remain weakly constrained, as expected given our limited redshift lever arm ($z \lesssim 2.3$). This residual flexibility in the dark‑energy sector motivates us to quantify whether different dark‑energy model assumptions can introduce measurable systematic biases in tests of gravity and cosmic curvature.

As shown in Table \ref{tab:wei_comparison}, compared with the two-probe constraint $\gamma_{\rm PPN}=1.11_{-0.09}^{+0.11}$ reported by Wei et al. (2022), the inclusion of BAO and CC into the SGL+SNe Ia baseline reduces the averaged 68\% uncertainty on $\gamma_{\rm PPN}$ by $\sim30\%$, demonstrating the strong complementarity of these cosmological probes. Therefore, our measurement of $\gamma_{\rm PPN}$ ranks among the more rigorous model-independent tests of general relativity on galactic scales. Meanwhile, the improvement is substantially more pronounced for $\Omega_k$, whose averaged 68\% uncertainty decreases by $\sim90\%$.

\begin{table}[htb!]
\centering
\renewcommand{\arraystretch}{1.4}
\caption{Quantitative comparison of model-independent constraints on $\gamma_{\rm PPN}$ and $\Omega_k$ between this work and the baseline result of Wei et al. (2022). The last row reports the reduction of the averaged lower and upper 68\% errors.} 
\label{tab:wei_comparison}
\begin{tabular}{l c c}
\toprule
\hline    
\hline
Work & $\gamma_{\rm PPN}$ (68\% C.I.) & $\Omega_k$ (68\% C.I.) \\
\midrule
\hline
Wei et al. (2022) (SGL+SNe Ia) & $1.11_{-0.09}^{+0.11}$ & $0.48_{-0.71}^{+1.09}$ \\
This work (SGL+SNe Ia+BAO+CC) & $1.119_{-0.072}^{+0.072}$ & $0.101_{-0.076}^{+0.077}$ \\
\hline
Averaged 68\% error reduction & $\sim 30\%$ & $\sim 90\%$ \\
\bottomrule
\hline
\end{tabular}
\end{table}

Beyond the two primary
parameters discussed above, the remaining background parameters are
constrained to
$H_0=68.903^{+0.877}_{-0.879}\ \mathrm{km\,s^{-1}\,Mpc^{-1}}$ and
$\Omega_m=0.285^{+0.022}_{-0.022}$, in agreement with standard
late-Universe determinations, whereas the two Chebyshev coefficients
$c_1=-0.217^{+0.301}_{-0.333}$ and
$c_2=-0.052^{+0.099}_{-0.105}$ are both consistent with zero within
$1\sigma$, showing no evidence for evolving dark energy over the
probed redshift range. The lens-population nuisance parameters of the
density-slope relation,
$\alpha_0=1.023^{+0.113}_{-0.121}$,
$\alpha_z=-0.239^{+0.076}_{-0.076}$, and
$\alpha_s=0.781^{+0.076}_{-0.072}$, together with the SN~Ia absolute
magnitude $M_B=-19.380^{+0.020}_{-0.021}$, take values consistent with
previous strong-lensing and Pantheon analyses and remain virtually
unchanged across the eight model-dependent fits discussed below. Having
established our fiducial model-independent results, we proceed in the
following subsection to quantify the systematic biases introduced by
adopting specific dark-energy model assumptions.

\subsection{Dark-energy-dependent Constraints on
$\gamma_{\rm PPN}$ and $\Omega_k$}
\label{subsec:ez_dependent}
\label{subsec:bias_calibration}
Motivated by the residual flexibility in the dark‑energy sector identified in the baseline analysis, we test whether the gravity inference is sensitive to the assumed dark‑energy background by repeating the joint four‑probe analysis, replacing the Chebyshev‑expanded dark‑energy $f_{\rm de}(z)$ in Eq.~\eqref{eq:E2_general1} in turn with the exact analytic dark‑energy factor $\mathcal{F}_{\rm de}(z)$ of eight representative models~\cite{magana2017testing}. For each model, we use its exact analytic form directly, rather than any polynomial approximation, to ensure an accurate and fair comparison.
All nine parallel fits (one model‑independent baseline plus eight model‑dependent fits) share identical likelihood formalism, observational datasets, and prior specifications as constructed in Sec.~\ref{subsec:stats}, differing only in the functional form of the dark‑energy evolution term. Each model‑dependent fit is obtained from Eq.~\eqref{eq:E2_general1} by the single substitution 
$f_{de}(z)\rightarrow\mathcal{F}_{\rm de}(z)$. The equation of state 
$w(z)$ and the corresponding exact factor 
$\mathcal{F}_{\rm de}(z)$ for each model are listed in Table~\ref{tab:de_models}, numbered consistently with Table~\ref{tab:all_model_constraints}. Models 1–5 are classical benchmark parameterizations of the dark-energy equation of state; Models 6–7 are constructed to remain regular in the formal limit $z\to-1$, at which the CPL parameterization diverges; Model 8 represents a physically motivated unified-dark-fluid scenario.

\begin{table*}[htb!]
\centering
\renewcommand{\arraystretch}{2.4}
\setlength{\tabcolsep}{6pt}
\caption{Equations of state $w(z)$ and exact dark-energy factors
$\mathcal{F}_{\rm de}(z)$ for the eight
models used in the parallel fits.}
\label{tab:de_models}
\begin{tabular}{c l >{$\displaystyle}l<{$} >{$\displaystyle}l<{$}}
\toprule
\hline
\hline
No & Name & w(z) & \mathcal{F}_{\rm de}(z) \\
\midrule
\hline
\textbf{1} & \textbf{$\Lambda$CDM}\,\cite{Barua:2025pdu} & -1 & 1 \\
\hline
\textbf{2} & \textbf{wCDM}\,\cite{wu2025comparison}
& w\;(\mathrm{constant}) & (1+z)^{3(1+w)} \\
\hline
\textbf{3} & \textbf{CPL}\,\cite{scherrer2015mapping,chevallier2001accelerating,linder2003exploring}
& w_0+w_a\dfrac{z}{1+z}
& (1+z)^{3(1+w_0+w_a)}
  e^{-\dfrac{3w_a z}{1+z}} \\
\hline
\textbf{4} & \textbf{JBP}\, \cite{magana2017testing}
& w_0+w_1\dfrac{z}{(1+z)^2}
& (1+z)^{3(1+w_0)}
  e^{\dfrac{3w_1 z^2}{2(1+z)^2}} \\
\hline
\textbf{5} & \textbf{BA}\,\cite{barboza2008parametric}
& w_0+w_1\dfrac{z(1+z)}{1+z^2}
& (1+z)^{3(1+w_0)}(1+z^2)^{\frac{3w_1}{2}} \\
\hline
\textbf{6} & \textbf{FSLL I}\,\cite{feng2012parameterization}
& w_0+w_1\dfrac{z}{1+z^2}
& (1+z)^{3(1+w_0)}e^{\frac{3w_1}{2}\arctan (z)}
  (1+z^2)^{\frac{3w_1}{4}}(1+z)^{-\frac{3w_1}{2}} \\
\hline
\textbf{7} & \textbf{FSLL II}\,\cite{feng2012parameterization}
& w_0+w_1\dfrac{z^2}{1+z^2}
& (1+z)^{3(1+w_0)}e^{-\frac{3w_1}{2}\arctan (z)}
  (1+z^2)^{\frac{3w_1}{4}}(1+z)^{\frac{3w_1}{2}} \\
\hline
\textbf{8} & \textbf{GCG}\,\cite{xu2012chaplygin,bento2002gchaplygin}
& -\dfrac{B_s}{B_s+(1-B_s)(1+z)^{3(1+\alpha_1)}}
& \left[B_s+(1-B_s)(1+z)^{3(1+\alpha_1)}\right]^{\frac{1}{1+\alpha_1}} \\
\hline
\bottomrule
\end{tabular}
\end{table*}

To quantify the amplitude and statistical significance of the shifts
in $\gamma_{\rm PPN}$ relative to the model-independent baseline, we
evaluate four diagnostic metrics for each parameterization:
\begin{itemize}
    \item \textit{Absolute offset}: the deviation between the
    model-dependent median and the baseline value,
    $\Delta\gamma = \gamma_{\rm model} - \gamma_{\rm baseline}$;
    \item \textit{Fractional offset}: the relative deviation normalized
    by the baseline median, $\Delta\gamma / \gamma_{\rm baseline}$;
    \item \textit{Statistical significance}: the absolute offset scaled
    by the baseline $1\sigma$ uncertainty,
    $|\Delta\gamma| / \sigma_{\gamma,\rm baseline}$;
    \item \textit{Error contraction ratio}: the ratio of the
    model-dependent uncertainty to the baseline uncertainty,
    $\sigma_{\gamma,\rm model} / \sigma_{\gamma,\rm baseline}$, which
    quantifies the degree to which a model prior may artificially
    narrow the inferred constraint.
\end{itemize}

Across all eight dark-energy models examined, we find highly
consistent median values and 68\% credible intervals for
$\gamma_{\rm PPN}$: the medians span the narrow range
$1.117$--$1.121$ with $1\sigma$ uncertainties of
$\sim 0.071$--$0.074$, in excellent agreement with the
model-independent baseline
$\gamma_{\rm PPN} = 1.119_{-0.072}^{+0.072}$. The model-dependent
offsets satisfy $-0.002\leq\Delta\gamma\leq+0.002$, corresponding to
fractional offsets $|\Delta\gamma|/\gamma_{\rm baseline}\lesssim0.2\%$,
and the largest absolute offset amounts to
$|\Delta\gamma|/0.072\simeq0.028\sigma$. The individual lower and
upper error half-widths relative to the baseline span
$0.986$--$1.028$. The absolute offsets are therefore uniformly
negligible, and no statistically meaningful error contraction or
expansion is induced by the model priors. 
The joint posterior constraints for all cosmological and nuisance parameters across the model-independent baseline and the eight dark-energy scenarios are compiled in Table~\ref{tab:all_model_constraints}, with the first row (Gen. Form) serving as the unbiased reference for column-by-column comparison with the remaining eight rows.

\begin{table*}[htb!]
  \centering
  \footnotesize
  \renewcommand{\arraystretch}{1.6}
  \caption{Comparison of MCMC joint constraints for the
  model-independent general form and eight dynamical dark-energy
  models (median $_{-1\sigma}^{+1\sigma}$).}
  \label{tab:all_model_constraints}
  \begin{tabular}{l *{10}{c}}
    \toprule
    \hline
    \hline
    Model &
    $\gamma_{\rm PPN}$ &
    $\Omega_k$ &
    $H_0$ &
    DE Param 1 &
    DE Param 2 &
    $\Omega_m$ &
    $\alpha_0$ &
    $\alpha_z$ &
    $\alpha_s$ &
    $M_B$ \\
    \midrule
    \hline
    \textbf{0. Gen. Form} &
    $1.119_{-0.072}^{+0.072}$ &
    $0.101_{-0.076}^{+0.077}$ &
    $68.903_{-0.879}^{+0.877}$ &
    $c_1=-0.217_{-0.333}^{+0.301}$ &
    $c_2=-0.052_{-0.105}^{+0.099}$ &
    $0.285_{-0.022}^{+0.022}$ &
    $1.023_{-0.121}^{+0.113}$ &
    $-0.239_{-0.076}^{+0.076}$ &
    $0.781_{-0.072}^{+0.076}$ &
    $-19.380_{-0.021}^{+0.020}$ \\
    \textbf{1. $\Lambda$CDM} &
    $1.117_{-0.071}^{+0.072}$ &
    $0.052_{-0.056}^{+0.058}$ &
    $68.807_{-0.802}^{+0.795}$ &
    $\cdots$ &
    $\cdots$ &
    $0.286_{-0.018}^{+0.018}$ &
    $1.022_{-0.119}^{+0.114}$ &
    $-0.234_{-0.074}^{+0.075}$ &
    $0.782_{-0.074}^{+0.074}$ &
    $-19.382_{-0.019}^{+0.019}$ \\
    \textbf{2. wCDM} &
    $1.119_{-0.073}^{+0.072}$ &
    $0.089_{-0.074}^{+0.076}$ &
    $68.931_{-0.868}^{+0.891}$ &
    $w=-1.055_{-0.096}^{+0.083}$ &
    $\cdots$ &
    $0.281_{-0.019}^{+0.019}$ &
    $1.024_{-0.123}^{+0.113}$ &
    $-0.238_{-0.076}^{+0.074}$ &
    $0.781_{-0.071}^{+0.076}$ &
    $-19.382_{-0.019}^{+0.020}$ \\
    \textbf{3. CPL} &
    $1.118_{-0.072}^{+0.073}$ &
    $0.101_{-0.074}^{+0.075}$ &
    $68.920_{-0.851}^{+0.850}$ &
    $w_0=-1.014_{-0.123}^{+0.107}$ &
    $w_a=-0.184_{-0.790}^{+1.024}$ &
    $0.276_{-0.038}^{+0.026}$ &
    $1.021_{-0.123}^{+0.116}$ &
    $-0.237_{-0.077}^{+0.073}$ &
    $0.782_{-0.073}^{+0.076}$ &
    $-19.381_{-0.020}^{+0.020}$ \\
    \textbf{4. JBP} &
    $1.118_{-0.074}^{+0.073}$ &
    $0.103_{-0.077}^{+0.077}$ &
    $68.876_{-0.847}^{+0.858}$ &
    $w_0=-1.012_{-0.167}^{+0.162}$ &
    $w_1=-0.489_{-1.453}^{+1.319}$ &
    $0.281_{-0.021}^{+0.022}$ &
    $1.019_{-0.123}^{+0.116}$ &
    $-0.240_{-0.078}^{+0.078}$ &
    $0.784_{-0.074}^{+0.077}$ &
    $-19.380_{-0.020}^{+0.020}$ \\
    \textbf{5. BA} &
    $1.118_{-0.071}^{+0.073}$ &
    $0.111_{-0.076}^{+0.077}$ &
    $68.883_{-0.871}^{+0.867}$ &
    $w_0=-1.009_{-0.113}^{+0.104}$ &
    $w_1=-0.293_{-0.511}^{+0.490}$ &
    $0.283_{-0.028}^{+0.024}$ &
    $1.020_{-0.122}^{+0.115}$ &
    $-0.239_{-0.075}^{+0.076}$ &
    $0.783_{-0.073}^{+0.077}$ &
    $-19.379_{-0.020}^{+0.020}$ \\
    \textbf{6. FSLL I} &
    $1.120_{-0.073}^{+0.072}$ &
    $0.107_{-0.076}^{+0.075}$ &
    $68.865_{-0.851}^{+0.873}$ &
    $w_0=-1.008_{-0.131}^{+0.119}$ &
    $w_1=-0.404_{-0.710}^{+0.647}$ &
    $0.283_{-0.021}^{+0.021}$ &
    $1.022_{-0.121}^{+0.113}$ &
    $-0.239_{-0.077}^{+0.074}$ &
    $0.782_{-0.072}^{+0.075}$ &
    $-19.379_{-0.020}^{+0.020}$ \\
    \textbf{7. FSLL II} &
    $1.121_{-0.072}^{+0.074}$ &
    $0.109_{-0.077}^{+0.077}$ &
    $68.908_{-0.852}^{+0.851}$ &
    $w_0=-1.033_{-0.103}^{+0.097}$ &
    $w_1=-0.825_{-1.201}^{+1.345}$ &
    $0.284_{-0.037}^{+0.026}$ &
    $1.025_{-0.123}^{+0.117}$ &
    $-0.240_{-0.077}^{+0.077}$ &
    $0.779_{-0.074}^{+0.079}$ &
    $-19.378_{-0.019}^{+0.019}$ \\
    \textbf{8. GCG} &
    $1.119_{-0.072}^{+0.072}$ &
    $0.105_{-0.065}^{+0.068}$ &
    $68.986_{-0.830}^{+0.841}$ &
    $B_s=0.902_{-0.101}^{+0.071}$ &
    $\alpha_1=0.245_{-0.165}^{+0.275}$ &
    $0.165_{-0.095}^{+0.074}$ &
    $1.022_{-0.118}^{+0.114}$ &
    $-0.240_{-0.075}^{+0.074}$ &
    $0.783_{-0.072}^{+0.075}$ &
    $-19.380_{-0.019}^{+0.019}$ \\
    \hline
    \bottomrule
  \end{tabular}
\end{table*}

Two conclusions emerge. First, none of the eight dark-energy scenarios produces a statistically significant shift in the inferred $\gamma_{\rm PPN}$: the offsets remain numerically negligible relative to the baseline $1\sigma$ width for both the standard 
$w(z)$-family parameterizations and the unified dark-fluid GCG scenario, and no model-dependent systematic bias is detected. Second, the constraint width is likewise insensitive to the model choice: all error-contraction ratios remain close to unity, and the lens-population nuisance parameters $(\alpha_0,\alpha_z,\alpha_s)$ are virtually unchanged across all nine fits. Because 
$\gamma_{\rm PPN}$ enters the joint likelihood solely through the SGL distance ratio, its residual uncertainty is governed by astrophysical systematics of the lensing sample rather than by the assumed expansion history; consequently, no cosmological model prior can artificially tighten the gravity test at the current level of observational accuracy. This robustness across all eight dark-energy scenarios points to a deeper structural origin rather than a numerical coincidence. Next, we analyze this origin through an order-by-order expansion of the exact dark-energy evolution factors in Section \ref{sec:discussion}, which reveals why the geometric probes at $z\lesssim 2.3$ cannot distinguish among these models.

\subsection{Analysis of the Geometric Origin of Dark-Energy Model Degeneracy}
\label{sec:discussion}
The negligible $\gamma_{\rm PPN}$ offsets reported in Section~\ref{subsec:ez_dependent} are not a numerical coincidence but reflect the intrinsic structure of the exact dark-energy factors. To reveal this structure, we expand each model’s exact factor $\mathcal{F}_{\rm de}(z)$—the replacement for the Chebyshev factor $f_{\rm de}(z)$ in Eq.~\eqref{eq:E2_general1}—in the compressed variable $u=z/(1+z)$ , which maps the observed redshift range onto $u\in[0,u_{\rm max}]$. Note that we adopt a Taylor rather than a Chebyshev basis for this expansion: for most dark energy models considered in this work, following the nonlinear mapping from $z$ to the compressed variable $u$, $\mathcal{F}_{\rm de}(u)$ is a transcendental function whose Chebyshev expansion coefficients must be evaluated via numerical integration and admit no closed-form expressions, whereas Taylor coefficients can be derived analytically order by order in a common basis, enabling direct term-by-term comparison across models.

The expansion coefficients up to third order, summarized in Table~\ref{tab:model_expansions}, exhibit a clear hierarchical degeneracy:
\begin{itemize}
    \item \textit{Zeroth order}: all eight factors are exactly degenerate, satisfying the normalization condition $\mathcal{F}_{\rm de}(0)=1$.

\item \textit{First order}: models 2–7 share the universal first-order coefficient $a_1=3(1+w_0)$ (reducing to $a_1=0$ for 
$\Lambda$ with $w_0=-1$), so 
$w_a$ and $w_1$ first contribute at second order. The GCG unified fluid, by contrast, exhibits a distinct first-order coefficient 
$a_1=3(1-B_s)$, but its impact on the distance–redshift relation remains small.
    \item \textit{Second and third orders}: differences within the $w(z)$ model family are minute, lying below the resolving power of geometric probes at $z\lesssim2.3$.
\end{itemize}

\begin{table*}[htb!]
  \centering
  \renewcommand{\arraystretch}{2.4}
  \setlength{\tabcolsep}{6pt}
  \caption{Third-order Taylor coefficients $a_i$ of each model's
  exact dark-energy factor
  $\mathcal{F}_{\rm de}(u)=\sum_{i=0}^{n}a_i u^{i}$, with
  $u=z/(1+z)$; $\mathcal{F}_{\rm de}(u)$ is the exact factor that
  replaces the Chebyshev $f_{\rm de}(z)$ in
  Eq.~\eqref{eq:E2_general1} for the parallel model fits.}
  \label{tab:model_expansions}
  \begin{tabular}{c l >{$\displaystyle}l<{$} >{$\displaystyle}l<{$} >{$\displaystyle}l<{$} >{$\displaystyle}l<{$} >{$\displaystyle}l<{$}}
    \toprule
    \hline
    \hline
    No & Name & 0th & 1st & 2nd & 3rd & $\dots$ \\
    \midrule
    \hline
    \textbf{1} & \textbf{$\Lambda$CDM} & 1 & 0 & 0 & 0 & $\dots$ \\
    \hline
    \textbf{2} & \textbf{wCDM} & 1
    & 3\left(1 + w\right)
    & \frac{3}{2} \left(4 + 7w + 3w^2\right)
    & \frac{1}{2} \left(20 + 47 w + 36 w^2 + 9 w^3\right)
    & $\dots$ \\
    \hline
    \textbf{3} & \textbf{CPL} & 1
    & 3\left(1 + w_0\right)
    & \frac{3}{2} \left(4 + 7w_0 + 3w_0^2 + w_a\right)
    & \frac{1}{2} \left[20 + 36 w_0^2 + 9 w_0^3 + 11 w_a + w_0 (47 + 9 w_a)\right]
    & $\dots$ \\
    \hline
    \textbf{4} & \textbf{JBP} & 1
    & 3\left(1 + w_0\right)
    & \frac{3}{2} \left(4 + 7w_0 + 3w_0^2 + w_1\right)
    & \frac{1}{2} (1 + w_0) \left(20 + 27 w_0 + 9 w_0^2 + 9 w_1\right)
    & $\dots$ \\
    \hline
    \textbf{5} & \textbf{BA} & 1
    & 3\left(1 + w_0\right)
    & \frac{3}{2} \left(4 + 7w_0 + 3w_0^2 + w_1\right)
    & \frac{1}{2} (5 + 3 w_0) \left(4 + 7 w_0 + 3 w_0^2 + 3 w_1\right)
    & $\dots$ \\
    \hline
    \textbf{6} & \textbf{FSLL I} & 1
    & 3\left(1 + w_0\right)
    & \frac{3}{2} \left(4 + 7w_0 + 3w_0^2 + w_1\right)
    & \frac{1}{2} \left[20 + 36 w_0^2 + 9 w_0^3 + 13 w_1 + w_0 (47 + 9 w_1)\right]
    & $\dots$ \\
    \hline
    \textbf{7} & \textbf{FSLL II} & 1
    & 3\left(1 + w_0\right)
    & \frac{3}{2} \left(4 + 7w_0 + 3w_0^2\right)
    & 10 + \frac{47 w_0}{2} + 18 w_0^2 + \frac{9 w_0^3}{2} + w_1
    & $\dots$ \\
    \hline
    \textbf{8} & \textbf{GCG} & 1
    & 3 \left(1 - B_s\right)
    & \frac{3}{2} \left(1 - B_s\right) \left(4 + 3 B_s \alpha_1\right)
    & \frac{1}{2} \left(1 - B_s\right) \left[20 + 9 B_s \alpha_1 (2 - \alpha_1) + 9 B_s^2 \alpha_1 (1 + 2 \alpha_1)\right]
    & $\dots$ \\
    \hline
    \bottomrule
  \end{tabular}
\end{table*}

These Taylor expansion coefficients of dark-energy evolution factor can provide a quantitative explanation for the negligible model-induced bias: nearly identical low-order coefficients produce nearly indistinguishable distance–redshift relations over the probed range, and hence statistically indistinguishable $\gamma_{\rm PPN}$ constraints. Geometric probes at $z\lesssim2.3$ are sensitive only to the first two or three orders of the $u$-expansion, whereas model-discriminating features reside predominantly at higher orders. This represents an intrinsic structural information ceiling of low-redshift distance data: unlike statistical uncertainties, it cannot be overcome by larger low-redshift samples or smaller measurement errors—contrary to the expectation that improved precision would enable model discrimination. This ceiling accounts for both the long-standing difficulty of distinguishing dynamical dark-energy models with geometric probes and the negligible bias in our gravity test. Overcoming it requires either distance measurements extending to the optimal dark-energy-sensitive redshift window or non-geometric probes sensitive to cosmic structure growth.

Two critical physical implications follow: (i) the near-degeneracy of low-order coefficients validates the conventional practice of computing strong-lensing distances within a standard $\Lambda$CDM background at current observational precision. Moreover, the residual $\gamma_{\rm PPN}$ uncertainty is relatively insensitive to the assumed expansion history: across all models considered, the choice of model prior alters neither the central value nor appreciably the width of the $\gamma_{\rm PPN}$ constraint; (ii) the difficulty in distinguishing among dynamical dark-energy models stems from limited information in low-redshift geometric measurements, not from sample size alone. Since model-discriminating terms appear predominantly beyond third order in the $u$-expansion, extending redshift coverage and incorporating structure-growth observables are inherently more effective than further enlarging low-redshift samples.

\section{Conclusions}
\label{sec:conclusion}
Galaxy-scale SGL systems, together with stellar velocity dispersion measurements of the lensing galaxies, offer a powerful means of testing the validity of GR by constraining the PPN parameter $\gamma_{\rm PPN}$ on kiloparsec scales. To test GR in this manner, however, one must know the angular diameter distances between the observer, lens, and source. In conventional SGL-based gravity tests, these distances are computed assuming the standard $\Lambda$CDM cosmology, which introduces a circularity issue when testing GR, since $\Lambda$CDM itself is  grounded in GR. In this work, to circumvent the circularity, we have employed the DSR in the FLRW metric to constrain both $\gamma_{\rm PPN}$ and $\Omega_{k}$, independently of any specific cosmological model. Instead of parameterizing the comoving distance directly as a polynomial in redshift, we expanded the dark‑energy evolution factor in second‑order Chebyshev orthogonal polynomials. This methodology preserves $\Omega_k$ as an explicit, purely geometric free parameter throughout the analysis, with curvature entering the formalism exclusively through the DSR, thereby eliminating the double-counting of curvature inherent to direct polynomial expansions of $d(z)$. In addition, the minimax property of Chebyshev polynomials ensures uniformly high approximation accuracy across the full redshift range, remedying the degraded convergence at high redshifts that plagues standard redshift-polynomial parameterizations. In addition, we extended the original two‑probe (SGL + SNe Ia) framework with BAO and CC as complementary probes. The resulting four‑probe joint analysis substantially breaks the strong $\gamma_{\rm PPN}$–$\Omega_k$ degeneracy that dominated the two‑probe uncertainty budget and markedly improves constraints on both parameters.

By combining 135 well-selected SGL systems with the reconstructed distance function from 1048 data points of SNe Ia, supplemented by BAO and CC measurements, we have obtained simultaneous estimates of $\gamma_{\rm PPN}$ and $\Omega_{k}$, without imposing any specific assumptions regarding either the theory of gravity and the dark energy model. Our results show that $\gamma_{\rm PPN}=1.119_{-0.072}^{+0.072}$ and $\Omega_k=0.101_{-0.076}^{+0.077}$ at the 68\% confidence level, providing a high-precision model-independent constraint on the post-Newtonian parameter on galactic scales. The measured $\gamma_{\rm PPN}$ is consistent with the GR prediction of unity at the $1.7\sigma$ level. For cosmic curvature, our result favors a mildly open universe, with $\Omega_k=0$ disfavored at only $1.3\sigma$ and thus compatible with spatial flatness within the $2\sigma$ confidence interval. Relative to the pioneering two-probe (SGL + SNe Ia) measurement of Wei et al. (2022), the averaged 68\% uncertainties are reduced by $\sim30\%$ for $\gamma_{\rm PPN}$ and by $\sim90\%$ for $\Omega_k$.

Using our model-independent measurement as an unbiased baseline, we quantify the systematic shifts in 
$\gamma_{\rm PPN}$ induced by eight representative dark-energy priors, including the 
$w(z)$ family and the generalized Chaplygin gas (GCG) unified fluid. Across all models, the central offsets remain well below the 
$1\sigma$ statistical uncertainty of the baseline, with no significant model-dependent bias, thereby validating the conventional practice of computing strong-lensing distances within a standard cosmological background: at current precision, dark-energy model choice introduces no measurable systematic effect on the inferred gravitational constraints.

Order-by-order Taylor expansions of exact dark-energy factors show that the uniformity of constraints across models stems from an intrinsic geometric degeneracy in low-redshift distance measurements, with model-discriminating terms lying beyond the reach of geometric probes at $z\lesssim2.3$—a structural information ceiling that cannot be overcome by improved precision or larger samples. This ceiling simultaneously explains the negligible model-induced bias in our gravity test and the long-standing difficulty of distinguishing dynamical dark-energy models, implying that overcoming it requires either distance measurements extending to the optimal dark-energy-sensitive redshift window or non-geometric probes sensitive to cosmic structure growth, thereby providing a quantitative reference for the observational strategies of forthcoming wide-field surveys such as LSST, Euclid, and CSST.

\section*{Acknowledgments}

We gratefully acknowledge Chen et al. (2019) for making their strong gravitational lensing catalog publicly available, and the Pantheon Collaboration for the public release of the Pantheon Type Ia supernova dataset (Scolnic et al. 2018). We also thank the DESI Collaboration for the publicly available baryon acoustic oscillation measurements from Data Release 1 (Adame et al. 2025), and the cosmic chronometer dataset presented in Loubser (2025).

This research made use of the open-source Python packages \texttt{emcee}, \texttt{corner}, \texttt{numpy}, \texttt{scipy}, \texttt{pandas}, \texttt{numba}, and \texttt{matplotlib}.

This work is partially supported by the National Natural Science Foundation of China (grant Nos. 12565009), the industry-funded collaborative research project ``AI-Enabled Integrated Circuit VR Immersive Educational Resource Development and Application'' (Project No. HX2025163) funded by Guizhou Juyiming Technology Co., Ltd., and the 2026 University-level General Scientific Research Project of Kaili University ``Study on Electromagnetic Wave Absorption Properties of Silicon–Barium Mineral-Based SiC Composite Thin Films'' (Project No. 2026YB005).

\end{document}